\documentclass{article}
\usepackage{spconf,amsmath,graphicx, amsfonts}
\usepackage[colorlinks,linkcolor=blue]{hyperref}
\usepackage{subfigure}

\title{Domain adversarial learning for emotion recognition}
\name{Zheng Lian$^{1,3}$, Jianhua Tao$^{1,2,3}$, Bin Liu$^{1}$, Jian Huang$^{1,3}$}
\address{$^{1}$National Laboratory of Pattern Recognition, CASIA, Beijing, China\\
	$^{2}$CAS Center for Excellence in Brain Science and Intelligence Technology, Beijing, China\\
	$^{3}$School of Artificial Intelligence, University of Chinese Academy of Sciences, Beijing, China}
\begin{document}
%
\maketitle
\begin{abstract}
In practical applications for emotion recognition, users do not always exist in the training corpus. The mismatch between training speakers and testing speakers affects the performance of the trained model. To deal with this problem, we need our model to focus on emotion-related information, while ignoring the difference between speaker identities. In this paper, we look into the use of the domain adversarial neural network (DANN) to extract a common representation between different speakers. The primary task is to predict emotion labels. The secondary task is to learn a common representation where speaker identities can not be distinguished. By using the gradient reversal layer, the gradients coming from the secondary task are used to bring the representations for different speakers closer. To verify the effectiveness of the proposed method, we conduct experiments on the IEMOCAP database. Experimental results demonstrate that the proposed framework shows an absolute improvement of 3.48\% over state-of-the-art strategies. 

\end{abstract}
\section{Introduction}

In many typical applications for emotion recognition, users do not always exist in the training corpus. The mismatch between training speakers and testing speakers leads to a performance degradation of the trained models. Therefore, it is vital to build speaker-independent systems for emotion recognition.

To ensure the model is speaker-independent, prior works focus on the data split strategies \cite{poria2017context, hazarika2018conversational}. These methods ensure no speaker overlap in the training set and the testing set. For example, the IEMOCAP dataset \cite{busso2008iemocap} contains five sessions and each session has different actors. Hazarika et al. \cite{hazarika2018conversational} used utterances from the first four sessions for training and others for testing. However, it is unclear whether these methods can actually learn speaker-independent representations.

Furthermore, obtaining large amounts of the realistic data is currently challenging and expensive for emotion recognition. The publicly available datasets (such as IEMOCAP \cite{busso2008iemocap} and SEMAINE \cite{mckeown2011semaine}) have relatively small number of total utterances. Prior works utilize unsupervised learning approaches to deal with low-resource training samples. One common method is to train autoencoders \cite{poultney2007efficient}. Autoencoders work by converting original features into compressed representations, aiming to capture intrinsic structures of the data. However, it is unclear whether compressed representations preserve the emotion component of the input. In fact, prior works have found the emotion component can be lost after feature compression \cite{busso2007interrelation}.


To deal with these problems, we focus on the domain adversarial neural network (DANN) \cite{ganin2016domain} for emotion recognition. Commonly, the mismatch in data distribution between the train (source domain) and test (target domain) sets leads to a performance degradation \cite{busso2013toward}. DANN relies on adversarial training for domain adaption, aiming to reduce the mismatch between the source domain and the target domain. In this paper, we use DANN to reduce the mismatch between different speakers, thus ensuring that the model can learn speaker-independent representations. Furthermore, different from previous unsupervised learning methods \cite{poultney2007efficient}, DANN can extract useful information from the unlabeled data while retaining the useful information for emotion recognition.

In this paper, we present a DANN based framework for emotion recognition. The main contributions of this paper lie in three aspects: 1) to deal with low-resource training samples, the proposed method can extract useful information from the unlabeled data while retaining emotion-related information; 2) to reduce the mismatch between speakers, the proposed method ensures the model can learn speaker-independent representations; 3) our proposed method is superior to other currently advanced approaches for emotion recognition. To the best of our knowledge, it is the first time that DANN is used for emotion recognition. 

\begin{figure*}[h]
	\centering
	\includegraphics[width=0.95\linewidth]{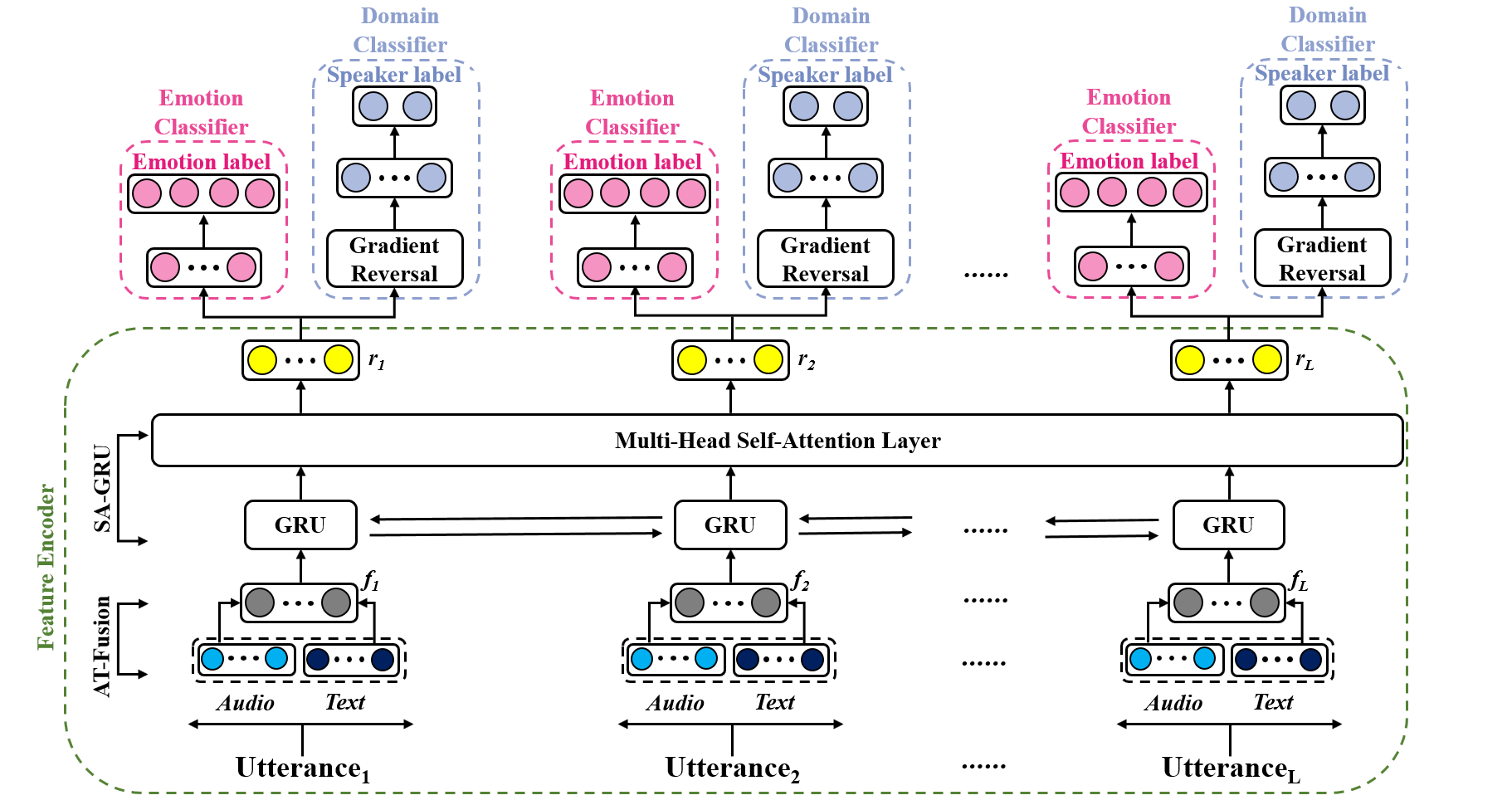}
	\caption{Overall structure of the proposed framework.}
\end{figure*}

\section{Proposed Method}
In this paper, we propose a multimodal learning framework for emotion recognition. As shown in Fig. 1, the proposed framework consists of three components: the feature encoder, the domain classifier and the emotion classifier.

\subsection{Problem Definition}
\textbf{Dataset:} We have a training set with annotated emotional data, and a testing set with unlabeled data. Like previous experimental settings in \cite{hazarika2018conversational}, speaker identities are available for both the training set and the testing set. 

\textbf{Task:} Define a conversation $U=[u_{1}, u_{2},..., u_{i},..., u_{L}]$, where $u_{i}$ is the $i^{th}$ utterance in the conversation and $L$ is the total number of utterances. The primary task is to predict the emotion labels of each utterance. The secondary task is to learn a common representation where speaker identities can not be distinguished.

\subsection{Feature Encoder}
The feature encoder contains two key components: the Audio-Text Fusion component (AT-Fusion) for multi-modalities fusion and the Self-Attention based Gated Recurrent Unit (SA-GRU) for contextual feature extraction.

\textbf{Multi-modalities Fusion (AT-Fusion):} Different modalities have different contributions in emotion recognition. To focus on the important modalities, we utilize the attention mechanism for multi-modalities fusion. Specifically, we first extract acoustic features and lexical features from each utterance. Then we equalize the dimensions of these features to size $d$ using two fully-connected layers, respectively. This provides the final acoustic features $a_{i} \in \mathbb{R}^{d \times 1}$ and lexical features $t_{i} \in \mathbb{R}^{d \times 1}$ for the utterance $u_{i}$. AT-Fusion takes $a_{i}$ and $t_{i}$ as inputs, and outputs the attention vector $\alpha_{fuse} \in \mathbb{R}^{1 \times 2}$ over these modalities. Finally, the fusion representation $f_{i} \in \mathbb{R}^{d \times 1}$ is generated as follows:
\begin{equation}
u_{i}^{cat} = Concat(a_{i}, t_{i})
\end{equation}
\begin{equation}
\alpha_{fuse} = softmax(w_F^T tanh(W_{F}u_{i}^{cat}))
\end{equation}
\begin{equation}
f_{i} = u_{i}^{cat}\alpha_{fuse}^T
\end{equation}
where $W_{F} \in \mathbb{R}^{d \times d}$ and $w_{F} \in \mathbb{R}^{d \times 1}$ are trainable parameters. Here, $softmax(x_i)=e^{x_i}/\sum_je^{x_j}$ and $u_{i}^{cat} \in \mathbb{R}^{d \times 2}$.

This multimodal representation is generated for utterances in the conversation $U$, marked as $F=[f_{1}, f_{2},...,f_{i},...,f_{L}]$.

\textbf{Contextual Feature Extraction (SA-GRU):} SA-GRU uses the bi-directional GRU (bi-GRU), in combination with the self-attention mechanism \cite{vaswani2017attention} to amplify the important contextual evidents for emotion recognition. Specifically, multimodal representations $F$ are given as inputs to the bi-GRU. Outputs of this layer form $H=[h_{1}, h_{2},...,h_{i},...,h_{L}]$, where $H \in \mathbb{R}^{L \times d}$. Then $H$ is fed into the self-attention network. It consists of a multi-head attention to extract the cross-position information. Each head $head_i \in \mathbb{R}^{L \times (d/h)}, i \in [1, h]$ ($h$ is the number of heads) is generated using the inner product as follows:
\begin{equation}
head_i = softmax((HW_i^Q)(HW_i^K)^T)((HW_i^V)
\end{equation}
where $W_i^Q \in \mathbb{R}^{d \times (d/h)}$, $W_i^K \in \mathbb{R}^{d \times (d/h)}$ and $W_i^V \in \mathbb{R}^{d \times (d/h)}$ are trainable parameters.

Then outputs of attention functions $head_i \in \mathbb{R}^{L \times (d/h)}, i \in [1, h]$ are concatenated together as final values $R \in \mathbb{R}^{L \times d}$. As contextual representations $R$ is generated for all utterances in the conversation $U$, it can also be represented as $R=[r_{1}, r_{2},...,r_{i},...,r_{L}]$, where $r_i \in \mathbb{R}^{d}, i \in [1, L]$.

\subsection{Domain Adversarial Neural Network for Emotion Recognition}
DANN is trained using labeled data from the training set and unlabeled data from the testing set. The network learns two classifiers -- the emotion classifier and the domain classifier. Both classifiers share the feature encoder that determines the representations of the data used for classification. The approach introduce a gradient reversal layer \cite{ganin2016domain} between the domain classifier and the feature encoder. This layer passes the data during forward propagation and inverts the sign of the gradient during backward propagation. Therefore, DANN attempts to minimize the emotion classification error and maximize the domain classification error. By considering these two goals, the model ensures a discriminative representation for the emotion recognition, while making the samples from different speakers indistinguishable.

In our proposed method, we train the emotion recognition task with the training set, for which we have emotion labels. For the domain classifier, we train the classifier with data from the training set and the testing set. Notice that the domain classifier does not require emotion labels, so we relay on unlabeled data from the testing set. These classifiers are trained in parallel. The objective function is defined as follows:
\begin{equation}
L = \frac{1}{n}\sum_{t=1}^{n}L_y^i - \lambda(\frac{1}{n+m}\sum_{t=1}^{n+m}L_d^i)
\end{equation}
where $L_y$ is the emotion recognition loss, $L_d$ is the domain classification loss, $n$ represents the number of labeled data from the training set, and $m$ represents the number of unlabeled data from the testing set. Here, $\lambda$ is a hyperparameter that controls the trade off between two losses.

Compared with the fully supervised learning strategy (where $\lambda=0$ in Eq. (5)), DANN has following advantages. Firstly, DANN learns a representation that confuses the domain classifier, which ensures the model speaker-independent. Secondly, DANN uses available unlabeled data to further reduce the mismatch between different speakers. These speaker-independent representations retain discriminative information learned during the training of the models with emotional data from the training set. Therefore, the proposed method can extract useful information from the unlabeled data while retaining discriminative information for emotion recognition.

\section{Experiments and Discussion}

\subsection{Corpus Description}
We perform experiments on the IEMOCAP dataset \cite{busso2008iemocap}. It contains audio-visual conversations spanning 12.46 hours of various dialogue scenarios. There are five sessions and a pair of speakers are grouped in a single session. All the conversations are split into small utterances, which are annotated using the following categories: anger, happiness, sadness, neutral, excitement, frustration, fear, surprise and other. To compare our method with state-of-the-art methods \cite{poria2017context, hazarika2018conversational}, we consider the first four categories, where happy and excited categories are merged into the single happy category. Thus 5531 utterances are involved. The number of utterances and dialogues of each session are listed in Table 1.
\begin{table}[h]
	\centering
	\caption{The data distribution of the IEMOCAP dataset.}
	\begin{tabular}{l|c|c|c|c|c}
		\hline
		Session      & 1    & 2    & 3    & 4    & 5    \\ \hline
		No.utterance & 1085 & 1023 & 1151 & 1031 & 1241 \\ \hline
		No.dialogue  & 28   & 30   & 32   & 30   & 31   \\ \hline
	\end{tabular}
\end{table}

\subsection{Experimental Setup}
\textbf{Features:} We extract acoustic features using the openSMILE toolkit \cite{eyben2010opensmile}. Specifically, we use the Computational Paralinguistic Challenge (ComParE) feature set introduced by Schuller et al. \cite{schuller2013interspeech}. Totally, 6373-dimensional utterance-level acoustic features are extracted, including energy, spectral, MFCCs and their statistics; In the meantime, we use word embeddings to represent the lexical information. Specifically, we employ deep contextualized word representations using the language model ELMo \cite{peters2018deep}. Compared with previous word vectors \cite{mikolov2013efficient}, these representations have proven to capture syntax and semantics aspects as well as the diversity of the linguistic context of words \cite{peters2018deep}. To extract utterance-level lexical features, we calculate mean values of word representations in the utterance. Totally, 1024-dimensional utterance-level lexical features are extracted. 

\textbf{Settings:} AF-Fusion equals the feature dimensions of different modalities to size $d=100$. SA-GRU contains a bi-GRU layer (50 units for each GRU component) and a self-attention layer (100 dimensional states and 4 attention heads). We test different $\lambda$ in Eq. (5) and we find that $\lambda=1$ gains the best recognition performance. To optimize the parameters, we use the Adam optimization scheme with a learning rate of 0.0001. We train our models for at least 200 epochs with a batch size of 20. $L2$ regularization with the weight $0.00001$ is also utilized to alleviate over-fitting problems. In our experiments, each configuration is tested 20 times with various weight initializations. The weighted accuracy (WA) is chosen as our evaluation criterion.

\subsection{Classification Performance of the Proposed Method}
Two systems are evaluated in the experiments. In additional to the proposed system, one comparison systems are also implemented to verify the effectiveness of our proposed method:

(1) Our proposed system (\textbf{\emph{Our}}): It is our proposed framework. For the emotion classifier, we train the classifier with data from the training set, for which we have emotion labels. For the domain classifier, we train the classifier with data from the training set and the testing set. As the domain classifier does not require emotion labels, we relay on unlabeled data from the testing set.

(2) Comparison system 1 (\textbf{\emph{C1}}): It comes from the proposed method, but ignoring the domain classifier. Specifically, we only optimize the emotion classifier by setting the $\lambda$ in Eq. (5) to be 0.

Furthermore, to explore the impact of the amounts of labeled samples in the training set, five training settings are discussed, including \emph{TS\_1234}, \emph{TS\_123}, \emph{TS\_134}, \emph{TS\_234} and \emph{TS\_23}. These training settings follow the same naming way. For example, \emph{TS\_123} represents that the training data contains Session 1$\sim$3, while the testing data contains other sessions (Session 4$\sim$5). As Session 5 always belongs to the testing data under these settings, we evaluate the classification performance on Session 5. Experimental results of WA are listed in Table 2.

\begin{table}[h]
	\centering
	\caption{Experimental results of two systems under different training settings.}
	\begin{tabular}{l|c|c|c|c|c}
		\hline
		 			& TS\_1234       & TS\_123        & TS\_134        & TS\_234     	& TS\_23      		\\ \hline
		C1       	& 81.06          & 80.82          & 79.85          & 78.89      	& 77.60     		\\ \hline
		Our         & \textbf{81.14} & \textbf{82.68} & \textbf{82.27} & \textbf{82.43} & \textbf{81.39} 	\\ \hline
	\end{tabular}
\end{table}

To verify the effectiveness of the proposed method, we compare the performance of the proposed method and {\emph{C1}}. Experimental results in Table 2 demonstrate that our proposed method is superior to {\emph{C1}} in all cases. Compared with {\emph{C1}}, our proposed method can learn speaker-independent representations. It ensures our model to focus on emotion-related information, while ignoring the difference between speaker identities. Therefore, this method can achieve better performance on unseen speakers (in Session 5). Furthermore, compared with {\emph{C1}}, the proposed method can use unlabeled samples in the training process. This semi-supervised approach uses unlabeled samples to further reduce the mismatch between different speakers. Meanwhile, this approach retains discriminative information learned during the training of the models with emotional data. Therefore, our proposed method is more suitable for emotion recognition than {\emph{C1}}.

To show the impact of the amount of training samples, we compare the performance under different training settings. Experimental results in Table 2 demonstrate that when we reduce training samples, {\emph{C1}} has 0.2\%$\sim$3.5\% performance decrement. Without enough training samples, {\emph{C1}} faces the risk of over-fitting. Therefore, the recognition performance on the unseen data becomes worse. Interestingly, we notice that our proposed method gains 0.2\%$\sim$1.5\% performance improvement when we reduce training samples. Meanwhile, we compare the performance of the proposed method and {\emph{C1}}. We observe that the margin of improvement increases with small amounts of training samples. These phenomenons reveal that if we utilize unlabeled samples properly, we can even achieve better performance than fully supervised learning methods. Different from previous unsupervised learning methods \cite{poultney2007efficient}, the proposed method can extract useful information from the unlabeled data while retaining discriminative information for emotion recognition.

\subsection{Comparison to State-of-the-art Approaches}
To verify the effectiveness of the proposed method, we further compare our method with other currently advanced approaches. Experimental results of different methods are listed in Table 3.
\begin{table}[h]
	\centering
	\caption{The performance of state-of-the-art approaches and the proposed approach on the IEMOCAP database.}
	\begin{tabular}{lc}
		\hline
		Approaches				 									& WA (\%)	      \\ \hline
		Rozgi{\'c} et al. (2012) \cite{rozgic2012ensemble}   		& 67.40           \\
		Jin et al. (2015) \cite{jin2015speech}   					& 69.20           \\
		Poria et al. (2017) \cite{poria2017context}   				& 74.31           \\
		Li et al. (2018) \cite{li2018inferring}     				& 74.80           \\
		Hazarika et al. (2018) \cite{hazarika2018conversational}	& 77.62           \\
		Li et al. (2019) \cite{li2019towards}						& 79.20           \\
		Proposed method							      				& \textbf{82.68}  \\ \hline
	\end{tabular}
\end{table}

Compared with our proposed method, these approaches \cite{poria2017context, hazarika2018conversational, rozgic2012ensemble, jin2015speech, li2018inferring, li2019towards} also utilized acoustic features and lexical features for emotion recognition. Context-free systems \cite{rozgic2012ensemble, jin2015speech, li2018inferring, li2019towards} inferred emotions based on only the current utterance in conversations. While context-based networks \cite{poria2017context} utilized the LSTMs to capture contextual information from their surroundings. However, context-based networks \cite{poria2017context} suffered from incapability of capturing inter-speaker dependencies. To model the inter-speaker emotion influence, Hazarika et al. \cite{hazarika2018conversational} used memory networks to perform speaker-specific modeling.

Experimental results in Table 3 demonstrate the effectiveness of the proposed method. Our proposed method shows an absolute improvement of 3.48\% over state-of-the-art strategies. This serves as strong evidence that the domain adversarial neural network can yield a promising performance for emotion recognition.

\section{Conclusions}
In this paper, we present a DANN based approach for emotion recognition. Experimental results demonstrate that our method enables the model to focus on emotion-related information, while ignoring the difference between speaker identities. Interestingly, we notice that our proposed method gains performance improvement when we reduce training samples. It reveals that our method can utilize unlabeled samples properly. Due to above advantages, this novel framework is superior to state-of-the-art strategies for emotion recognition.


\bibliographystyle{IEEEbib}
\bibliography{mybib}

\end{document}